# Metrics sonification: The introduction of new ways to present bibliometric data using publication data of Loet Leydesdorff as an example

Lutz Bornmann*, Rouven Lazlo Haegner**

*Science Policy and Strategy Department

Administrative Headquarters of the Max Planck Society

Hofgartenstr. 8,

80539 Munich, Germany.

Email: bornmann@gv.mpg.de

**Butterama Recording Center

Ziegrastr. 11-13

12057 Berlin

Email: haegner@butterama.com

**Abstract**

The visualization of publication and citation data is popular in bibliometrics. Although less common, the representation of empirical data as sound is an alternative form of presentation (in other fields than bibliometrics). In this representation, the data are mapped into sound and listened to by an audience. Approaches for the sonification of data have been developed in many fields since decades. Since sonification has several advantages for the presentation of data, this study is intended to introduce sonification to bibliometrics named as 'metrics sonification'. Metrics sonification is defined as the sonification of bibliometric information (measurements, data or results) for their empirical analysis and/or presentation. In this study, we used metadata of publications by Loet Leydesdorff (named as Loet in the following) to sonify their properties. Loet was a giant in the field of scientometrics, who passed away in 2023. The track based on Loet's publications can be listened to on SoundCloud using the following link: https://on.soundcloud.com/oxBTA32x4EgwvKVz5. The track has been composed in F minor; this key was chosen to express the sad occasion. The quantitative part of the track includes a parameter mapping (a sonification) of three properties of his publications: (1) publication output, (2) open access publication, and (3) citation impact of publications. The qualitative part (spoken audio) focuses on explanations of the parameter mapping and descriptions of the mapped papers (based on their titles and abstracts). The sonification of Loet's publications presented in this study is only one possible type of metrics sonification application. As the great number of projects from other disciplines have demonstrated, many other types of applications are possible in bibliometrics.

**Keywords**

bibliometrics, data sonification, musical sonification, data music, metrics sonification, scientometrics

# 1 Introduction

The visualization of results based on publication and citation data is prevalent in bibliometrics. One does not only find scatter plots, beam plots, and box plots in bibliometric papers, but also many maps of science that, e.g., visualize collaborations between institutions (Bornmann, Stefaner, de Moya Anegón, & Mutz, 2016) or developments of research fronts (Boyack & Klavans, 2010). An overview of the broad spectrum of available science maps, which are usually defined as “visual representations of the structure and dynamics of scholarly knowledge” (Petrovich, 2020), can be found in Petrovich (2020). The popularity of visualizations is not specific for the bibliometric domain; Dayé and de Campo (2006) denote the Western culture as a whole as a visual culture: “It is a culture of seeing, of reading, a culture of scripture and images. Science, as a part of this culture, has relied and continues to rely on the perceptual capacities of the eye” (p. 352). Although other modes of perceptions are possible, visualizations are the most popular form of presenting data or empirical results.

Although less common (Dayé & de Campo, 2006), the representation of measurements, data or empirical results as sound is an alternative form of presentation. In this representation, the measurement, data or empirical results are mapped into sound and are accordingly presented. An early example of data sonification is an experiment by Galileo Galilei “with a ball falling on an inclined plane: A small heavy ball was released down an inclined plane so that as it rolled it lightly touched catgut strings that were tightened above the plane … Galileo noticed that every time he repeated the experiment, the sound of the strings had the same rhythm” (Dayé & de Campo, 2006, p. 352). Other more recent and well-known examples are the widely used Geiger counter and pulse oximeter. The Geiger counter sonifies the detection and measurement of ionizing radiation by crackling sounds (Supper, 2015). The Geiger counter immediately translates the measurement into sensory perception. The pulse oximeter is standard equipment in medical operations; it produces “a tone that varies in pitch with the level of oxygen in a patient's blood, allowing the doctor to monitor this critically important information while visually concentrating on surgical procedures" (Kramer et al., 2010, p. 12).

Since for certain empirical applications sonification is an interesting alternative or addition to visualization, approaches for the sonification of measurements, data or empirical results have been developed in many disciplines. According to Lindborg, Lenzi, and Chen (2023) sonification is an "emerging discipline that struggles to define its boundaries, its impacts and more importantly, shared methods, processes, and tools for the mapping of data to sound" (see also Walker & Nees, 2011). Many examples of sonification can be found in the Data Sonification Archive (see https://sonification.design) or have been published in the proceedings of the ICAD (International Community for Auditory Display) conference, which "is an annual international venue for sonification and other auditory display research" (Kramer et al., 2010, p. 16).

Since sonification has already been introduced in many fields dealing with empirical data but not in bibliometrics, this study is intended to catch up on this introduction based on a case study. We would like to name the sonification of bibliometric data 'metrics sonification'. Metrics sonification is defined thus as the sonification of bibliometric information (measurements, data or results) for their empirical analysis and/or presentation. For Dayé and de Campo (2006), especially data from the social sciences – and bibliometric data are social science data – are well suited for data sonification, because "they are multidimensional, and they usually depict complex relations and interdependencies" (Dayé & de Campo, 2006, p. 355).

We see several reasons why metrics sonification can develop to a relevant technique in bibliometrics. (1) One important reason is research results from other fields than bibliometrics that pointed out the usefulness of sonification to get an (initial) insight in empirical data (Walker & Nees, 2011) such as astronomical and space science. In this study, we demonstrate only one possible application type of sonification in bibliometrics allowing to have an insight in an empirical dataset, but many other application types are imaginable. This study may be an opener for forthcoming studies using different sonification approaches (used in other fields). (2) For around ten years, several reform movements for 'responsible research assessment' have emerged (Rushforth & de Rijcke, 2023). These movements focus on the problems of metrics (such as the problematic use of the Journal Impact Factor; see Pulverer, 2015) and the recommendation to foster expert judgement (Hicks, Wouters, Waltman, de Rijcke, & Rafols, 2015). We empirically demonstrate in this study that metrics sonification is able to combine (the presentation of) quantitative metrics and qualitative expert judgements. The musical track composed in this project with the case study data includes sonified bibliographic metadata as well as explanations of the research underlying the data. The specific advantage of sonification for building a bridge between qualitative and quantitative approaches has already been outlined and demonstrated by Agné, Sommerer, and Angeler (2020) for the field of political science.

(3) In post-academic science, it is usual to communicate (transfer) research results to the general public. Altmetrics (e.g., Twitter counts) have been specifically introduced in the area of scientometrics to measure whether the transfer has happened (Tahamtan & Bornmann, 2020): "In order for society to make effective policy decisions on complex and far-reaching subjects, such as appropriate responses to global climate change, scientists must effectively communicate complex results to the non-scientifically specialized public" (Larsen & Gilbert, 2013). Sonification has been proposed as a suitable technique to facilitate this communication. For example, Larsen and Gilbert (2013) developed "an approach by which music can be used as public outreach to engage the public in microbial ecology": the approach 'Microbial Bebop' transforms microbial environmental data into music. In scientometrics, only a few techniques exist to support the transfer of results from empirical research to a public audience. Impressive data visualizations may be an appropriate way such as the meaningful visualizations proposed by González-Márquez, Schmidt, Schmidt, Berens, and Kobak (2023). The sonification of bibliometric data (including aesthetic musical compositions) may be another way (following, e.g., Larsen & Gilbert, 2013). The term 'aesthetic' refers here to "the quality of unifying sensory experiences; of an event, an object, or an environment" (Worrall, 2019, p. 144). The aesthetic experiences of the metrics sonification should be attractive to the listener from the public; otherwise the interest in the bibliometric results would ease up.

The specific application of sonification in this study can be seen as a fourth motivation to sonify bibliometric data in similar contexts. Bibliometric data are well suited to describe the scientific career of eminent scientists (e.g., who passed away or received honorable prizes). Their research topics are usually covered in the titles and abstracts of the publications, the citation impact of these publications can be seen as a proxy for their influence on other scientists, and the cited references in the publications reflect the historical roots of their research. These are the reasons why Bornmann, Haunschild, and Leydesdorff (2018) used such data as input to a Reference Publication Year Spectroscopy (RPYS) for analyzing the historical roots of one of the founder of bibliometrics: Eugene Garfield. In this study, we used metadata of publications by Loet Leydesdorff (named as Loet in the following) to sonify their properties. Loet was a giant in the field of scientometrics; one of the authors of this paper (LB) published more than 70 papers with him in a close collaboration over many years (see here Bornmann & Haunschild, 2023). With this study, we do not only use the papers published by Loet as a suitable example for metrics sonification, we dedicate the introduction of metrics sonification to the field of bibliometrics to Loet.

After the definition of sonification and its history in Section 2 and a description of the dataset used for the sonification of Loet's papers in Section 3, a combined account of methods and results follows in Section 4. We decided to combine both methods and results in one section because of two reasons: (1) Since this paper introduces metrics sonification in bibliometrics, it seemed to be the best way for us to explain the method in close connection to the dataset used. (2) The music track (the sonification), i.e. the central finding of the current empirical project, cannot be presented in the results section of the paper (which differentiates sonification from most visualizations). Only the link to the platform with the track can be presented in the paper. Section 5 discusses the proposed sonification and the potential of sonification for the bibliometric field.

# 2 Definition of sonification and its history

Data sonification – or "data music" (Vickers, 2016) – is a subtype of auditory displays which can be "defined as any display that uses sound to communicate information" (Walker & Nees, 2011, p. 9). Many definitions of data sonification can be found in the literature (see, e.g., Ludovico & Presti, 2016; Supper, 2015; Worrall, 2019); most of the definitions are similar and refer to an initial definition of Kramer et al. (2010): "sonification is defined as the use of nonspeech audio to convey information. More specifically, sonification is the transformation of data relations into perceived relations in an acoustic signal for the purposes of facilitating communication or interpretation" (p. 4).

Sonifications where sound elements are tied to data (Zanella et al., 2022) can be denoted as the pendant to visualizations where graphical elements are tied to data. Similar to visualizations, sonifications can reflect categories or names, relative sizes of scores or exact values of data (Nasir & Roberts, 2007). An extended (compared to Kramer et al., 2010), recently proposed definition of sonification can be found in Liew and Lindborg (2020): sonification is "any technique that translates data into non-speech sound, with a systematic, describable, and reproducible method, in order to reveal or facilitate communication, interpretation, or discovery of meaning that is latent in the data, having a practical, artistic, or scientific purpose" (p. 25). The people listening to the sound can reason about the data, "interpret them, form hypotheses and thus add context, so that the represented data become informative. In sonification, we listen to data in order to gather information" (Dayé & de Campo, 2006, p. 354).

Sonification is an interdisciplinary undertaking that integrates "concepts from human perception, acoustics, design, the arts, and engineering" (Kramer et al., 2010, p. 4). Projects in sonification involve people from various disciplines such as "psychologists, computer scientists, engineers, physicists, composers, and musicians" (Kramer et al., 2010, p. 4). Further relevant disciplines for sonification are audio engineering, audiology, informatics, linguistics, mathematics, psychology, and telecommunications (Walker & Nees, 2011). Sonification is used in sonification projects to explore data, monitor complex processes or assist to navigate in data spaces (Rönnberg & Lenzi, 2022). Listening to data can support to "search out inherent structures, trends or patterns" (Dayé & de Campo, 2006, p. 350). Walker and Nees (2011) attribute – with reference to some previous publications – to data sonification three basic functions: (1) alerts, alarms, and warnings (the sound based on data stimulates immediate action in a certain context), (2) status and monitoring messages (the sound reflects a current or an ongoing status of a system); and (3) data exploration (the sound encodes information about a dataset). Metrics sonification as introduced in this study is connected especially to the last function: data exploration.

The use of sound as an alternative or addition to visualization in science started around the late 1980s (Lenzi & Ciuccarelli, 2020): "The availability of microprocessors with built-in sound cards and digital sound synthesis allowed for the creation of a large variety of different sounds based on a similarly large variety of data. Researchers in human-computer interaction, for instance, began to explore how graphical user interfaces could be accompanied by audio and how sound could help in the exploratory analysis of multivariate datasets" (Supper, 2015, p. 447). The International Conference on Auditory Display (ICAD, https://icad.org) was organized by Gregory Kramer in 1992 for the first time at the Santa Fe Institute (Santa Fe, NM, USA). For Supper (2015), the first ICAD was the starting point for building a community specifically dedicated to data sonification. The ICAD is the most important conference in the area of data sonification today.

Different approaches for data sonification have been proposed over the years. Summarizing classifications of the different approaches can be found in Walker and Nees (2011) and Worrall (2019). Walker and Nees (2011) classified the approaches as event-based sonifications (data dimensions are reflected in sound dimensions), model-based sonifications (the listener interacts with a virtual model, and the model's properties are data informed), and continuous sonifications (time-series data are constantly transformed into sound). Worrall (2019) presents another triple classification as Walker and Nees (2011) that, however, also includes continuous (data) sonifications. The other two are discrete data representations (each data point is connected to a corresponding sound event) and direct data representations (the data is directly translated into sound). All these classifications from both publications (with the exception of model-based sonifications) can be summarized under the term 'parameter mapping': "data dimensions are mapped to sound parameters: either to physical (frequency, amplitude), psychophysical (pitch, loudness) or perceptually coherent complexes (timbre, rhythm)" (Worrall, 2019, p. 38). So, it is not surprising that parameter mapping is "the most common technique of data sonification in use today" (Worrall, 2019, p. 39) and it is used also in this study.

What are the specific advantages of data sonification? Why do we need sonification besides visualization in science and beyond? Several advantages have been named in the literature: (1) It has been outlined that the human auditory system is superiorly able to recognize temporal patterns, changes and other highly complex properties of sound or information (Campo & Campo, 1999; Dayé & de Campo, 2006; Kramer et al., 2010; Walker & Nees, 2011). It has also been claimed that the ear is one of the best measuring instruments of the human body (Dayé, 2006). (2) Sonification allows more design alternatives than visualization, and it can thus be experienced with far more dimensions synchronously (Dayé, 2006): “Sound is inherently multi-dimensional because it is characterised by various parameters (e.g., pitch, volume, tempo, location in a stereo field, timbre)” (Zanella et al., 2022, p. 1241). (3) Data sonification allows the accessibility of scientific knowledge auditory to impaired (non-expert) audience (Lindborg et al., 2023; Zanella et al., 2022). (4) The combination of sound and visualization makes people “more effective at gaining an initial overview of the data and/or for identifying signals and features in astronomical and space science data” (Zanella et al., 2022, p. 1241). The increase of efficiency in the reception of information by applying two channels is also mentioned by Lenzi and Ciuccarelli (2020). (5) Enjoyment (listening to music) besides usefulness (exploring structures in data) is another point that speaks for sonification (Lenzi & Ciuccarelli, 2020).

The advantages of data sonification that have been named in the literature are also offset by several disadvantages: (1) Most people are not trained in acknowledging auditory patterns to such an extent as they are trained in acknowledging patterns in visualizations (Zanella et al., 2022). Thus, acknowledging auditory patterns and listening carefully need to be learned or trained, before sonification may develop to an established instrument (in bibliometrics) (Dayé & de Campo, 2006; Kramer et al., 2010). (2) There is the danger that the sonification of data does not support the interpretation and explanation of this data. This problem has been outlined for previous sonifications: “By focusing mostly on technical questions about how data can be made audible, the interpretation of the underlying data often slips into the background” (Supper, 2015, p. 448). To target this danger, specialists in the empirical domains should be involved in the sonification process. (3) Supper (2015) points out that sound decisions are subjective and may “therefore [be] unsuitable for scientific analysis” (p. 450). The sonification may appear “arbitrary to the listener in relation to the underlying data” (Lindborg et al., 2023). In many sonification projects, there seems to be uncertainty about the proper connection of information (data) and interpretation (sound) to produce meaningful sonifications. This problem has been named as ‘mapping problem’ (Lenzi & Ciuccarelli, 2020). The subjectivity of sound decisions and uncertainty in the use of sound could be tackled, however, by developing generally accepted standards similar to visualizations (Zanella et al., 2022). The development of these standards in sonifications may be more complicated than in data visualizations, since sound is (very) culture-dependent. In the development of standards, it should be considered therefore that "many symbolic aspects of sound are culture-dependent" (Ludovico & Presti, 2016, p. 73).

Since the beginning of sonification in the (late) 1980s, many projects in various disciplines have been carried out such as seismology, astronomy, geography, and cartography, as well as social sciences (Lenzi & Ciuccarelli, 2020): "Although some sonifications are couched in a relatively traditional musical idiom, such as the jazz of the Microbial Bebop project, many eschew traditional musical conventions in favor of more abstract sounds. The applications for which sonifications are developed range from analysis tools for scientific specialists to musical pieces; in fact, many projects playfully straddle the boundary between science and art" (Supper, 2015, p. 442). A recent overview of sonification projects in the area of climate change can be found in Lindborg et al. (2023). The investigation by Zanella et al. (2022) on sonification projects in astronomy reveals that most projects (more than 60%) combine sonification (using the parameter mapping technique) with visualizations with the objective of strengthening public engagement in research.

# 3 Dataset and sounds used

In the Web of Science (WoS) Core Collection (Clarivate), we found 414 papers in Loet's researcher profile which have been published between 1980 and 2023 (date of search, October 6, 2023). Since we are interested in this study to sonify – with the field-normalized citation impact of his papers and the Open Access (OA) status – specific data, we used the in-house version of the WoS by the Competence Network Bibliometrics (KB) that includes this data. The data used for this study also covers the abstracts of the papers, which were an important source of information to interpret the papers. Only papers with an at least three-year citation window have been considered to have a reliable impact measurement over a longer period (and therefore all papers were discarded that were published after 2019).

In this study, we used the mean normalized citation score (MNCS) to measure citation impact field-normalized. This indicator divides the citation counts of a focal paper by an expected citation score. The expected citation score is the mean citation impact of all papers which have been published in the same subject category and publication year as the focal paper. A MNCS of 1 means that the focal paper received an average impact; a MNCS of 1.4 means that the impact is 40 percentage points higher than the average. Loet's papers were also classified as Open Access (OA) papers or not: OA "literature is digital, online, free of charge, and free of most copyright and licensing restrictions. OA removes price barriers (subscriptions, licensing fees, pay-per-view fees) and permission barriers (most copyright and licensing restrictions)" (Suber, 2015). The access to non-OA papers is restricted (by a paywall).

The downloaded number of publications (n=296) with all relevant information from the KB in-house database is reduced compared to Loet's publication set in the WoS Core Collection (n=414). Either not all relevant information are available for the publications or the publications have appeared after 2019. The downloaded papers were sorted by publication year and month to have a dataset with time-series information for the sonification: "Sound is a time-bound phenomenon; it only exists in time, and it takes time to perceive sounds and their interrelations" (Dayé & de Campo, 2006, p. 355).

We included in the composed track several sounds. Some sounds are from Splice which is a cloud-based music creation platform (see https://splice.com). The platform is organized in packs including specific sounds (.wav files). Each sound has a specific function in the sonification of Loet's publications. The used (produced) sounds and their functions are as follows:

Background drone: The background drone consists of the piece “Minor Cadenza” by the composer Hennadii Boichenko. He composed the piece exclusively for our sonification against the backdrop of the sad occasion. The piece consists of synthesized string sounds in F minor, evoking a somber and atmospheric mood. The drone predominantly features a rich texture of sampled string sounds, played by Hennadii Boichenko himself on the piano. The drone’s pitch subtly changes throughout different parts of the track, creating a sense of progression and maintaining the listener’s interest. The drone is occasionally time-stretched or faded into subsequent sections to ensure smooth transitions.

Drums (kick, snare, hi-hat, and crash): Drums are played on several instances in the track by using the synthesizer Fabfilter Twin 3 Plugin. To create different sounds for kick, snare, and hi-hat, we modulated the oscillators of the synthesizer.

Bassline: The bassline is played on a Höfner 182 bass guitar from 1964. It is played through a simulation of the Fender Bassman amplifier. The direct signal from the bass guitar is mixed with the simulated bass amp signal.

Publication output: We used the sound ESM_Explainer_Video_One_Shot_Motion_Whoosh_Low_Air_Short_3.wav from Splice.

Publication event (OA): We used the sound KSHMR_Sub_Drop_03.wav from Splice.

The sonification presented in this study has been produced using the music production software Ableton Live (version 11) and Reaper (version 7). Reaper is an open source digital audio workstation from Cockos Incorporation (San Francisco, CA, USA).

# 4 Methods and Results

It is the objective of the sonification in this study to combine qualitative and quantitative evaluation approaches to assess and explain Loet’s publications. The quantitative part of the musical track includes a parameter mapping of three properties of his publications: (1) publication output, (2) OA publication, and (3) citation impact (MNCS) of a publication.

The qualitative part (spoken audio) focuses on explanations of the parameter mapping and descriptions of the mapped papers (based on their titles and abstracts). Since the sonification is intended to have a strong focus on the content of single publications, it was necessary to reduce Loet's publication set used for the sonification. The sonification was composed in four-four time; thus, we selected 32 publications from the early phase of Loet's career and 32 publications from the late phase (2x8x4 publications).

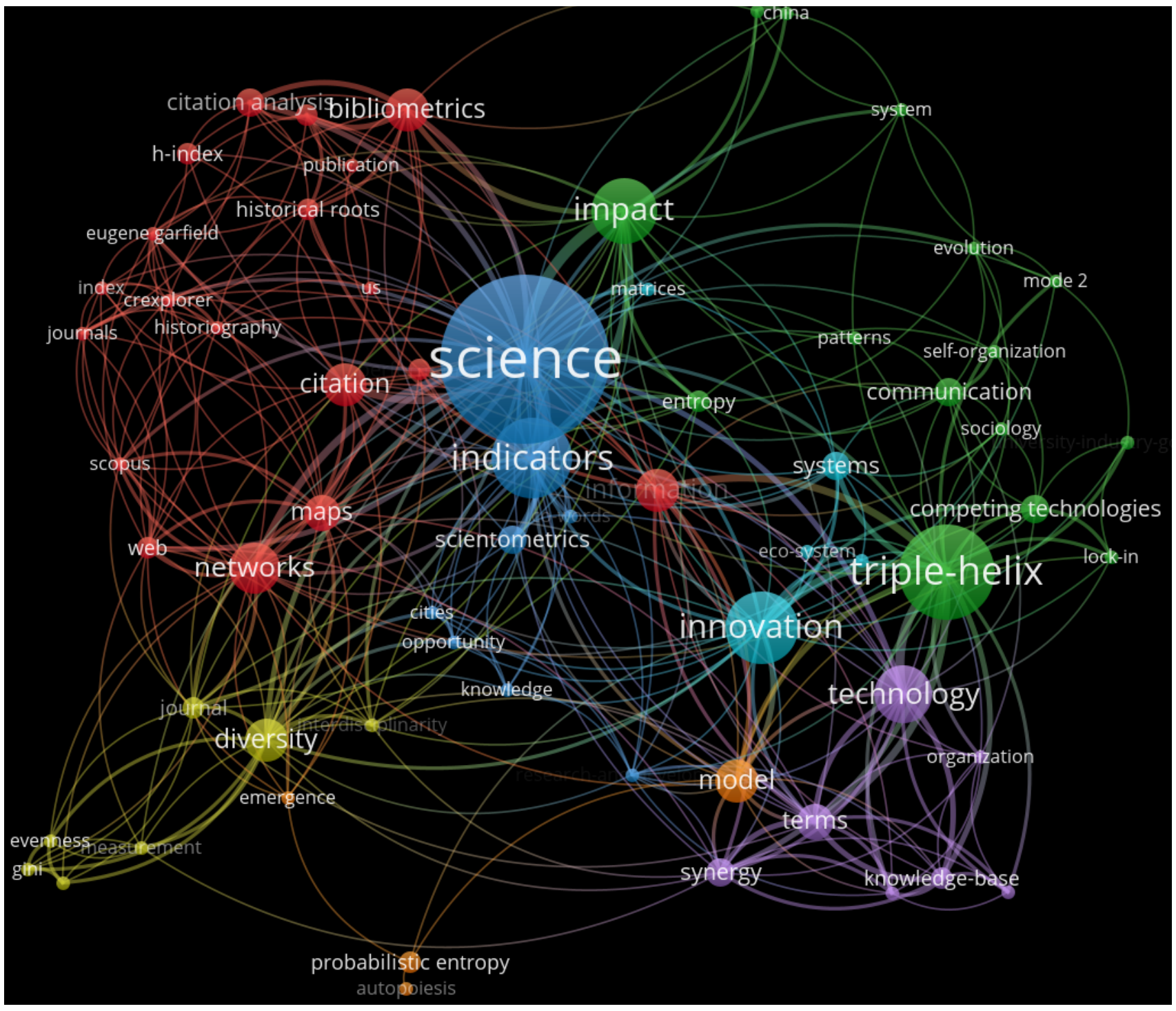

Figure 1. Keyword co-occurrence network of 64 papers published by Loet in the early and late career. The network is used as background image for the uploaded track on SoundCloud that is shown during the listening of the track.

The track composed sonifying Loet's publications can be listened to on SoundCloud (a popular audio streaming service) using the following link: https://on.soundcloud.com/oxBTA32x4EgwvKVz5. The track has around 10 minutes including the different parts with spoken audio (explained in Section 4.1) and parameter mapping (explained in Section 4.2). In order to compose a track that will be an aesthetic enjoyment for the listener and to convey a sad mood to the listener, we included parts with electronic music without a connection to publication data in the track (explained in Section 4.3).

When the user listens to the track (e.g., on the iOS platform) using SoundCloud, a background image appears during the listening (see Figure 1). SoundCloud provides this service for uploaded tracks. We decided to produce a background image with a close connection to the composed sonification. Since the sonification focuses on Loet's papers from the early and late phases of his career, the background image includes the following visualization produced with the VOSviewer program (van Eck & Waltman, 2010) – a program for constructing and visualizing bibliometric networks: a co-occurrence network of keywords from papers published in the early and late phases. It is the intention of keywords to reflect the content of a paper. The nodes in the network reflect the frequency of keyword occurrences in the publications from the phases. The links between the nodes in the networks are based on the co-occurrences of two keywords in the same publication; co-occurrences reflect therefore the number of publications in which two keywords occur together in keyword lists. By inspecting the network in Figure 1, the listener of the composed track receives additional information to Loet's publications (as visualization).

### 4.1 Spoken audio: the qualitative approach of sonification

In the literature on sonification research and application, different opinions on the inclusion of spoken audio in sonification can be found. For example, the above cited definition of sonification by Liew and Lindborg (2020) explicitly excludes spoken audio. In another paper, however, such exclusions are questioned: "it is hard to uphold that 'visualization' should exclude visual symbols such as text, numbers, emojis etc., at least in practice. The exclusion of 'speech sound' in sonification has been questioned" (Lindborg et al., 2023). The exclusion of spoken audio from sonification has been seen critically also by other authors (Worrall, 2019).

It is usual in scientific writing to explain all figures included in a manuscript in such a way (in the title and notes) that they are self-explanatory (see, e.g., American Psychological Association, 2020). By using spoken audio, it is possible to do the same with sonification: "it is often hard to understand quantities from sonifications, but speech provides the user with exact quantifiable information" (Nasir & Roberts, 2007, p. 117). In contrast to visualizations, which are embedded in scientific papers in most cases, sonifications are usually produced as outcomes that stand on their own including mapped parameters and their explanations. In order to reach the goal that the sonification of Loet's publications is self-explained, we included some spoken audio passages discussing the papers in the sonification. The spoken audio was produced based on text passages that were spoken by an English native speaker. The text passages have different purposes in the track:

At the beginning of the track, we start with a general introduction to the person 'Loet Leydesdorff' and the purpose, data, and structure of the sonification: "Professor Loet Leydesdorff from Amsterdam passed away in March 2023. He was a giant in his research area called the science of science studies. Only a few other researchers have had a similarly important contributing impact on this area. Starting in the 1980s, Loet published more than 400 papers until 2023, which are covered in the core collection of the Web of Science database. Most of his studies are based on scientometric methods targeting research questions from the history, philosophy, sociology, and economy of science. In memory of Loet, we composed this metrics sonification to sonify two phases of his career in research, the early and the late phase. The sonification of each phase has two parts. In the first part, selected publications from the phase are explained by spoken words. In the second part, certain properties of the papers are sonified by using the parameter mapping technique".

In parameter mapping of data, it is important that the listener understands how the sonification can be understood “so the listener can perform the interpretation tasks” (Walker & Nees, 2011, p. 26). In order to support the understanding of the parameter mapping in the composed track for Loet’s papers, we included the following explanation before the listener hears the part with the data sonification:

1. “This sound reflects a publication event whereby the publication appeared with open access or not. [The corresponding sound is played].
2. The following sound signals an open access publication. [The corresponding sound is played].
3. Citation impact appears as pitch variation of this medium citation impact tone. [The corresponding sound is played].
4. Outreaching citation impact is reflected by such a high tone. [The corresponding sound is played].
5. Such a low tone means far below citation impact. [The corresponding sound is played].
6. The citation impact class is additionally spoken by numbers between one and seven. [Some numbers are spoken]”.

   The class numbers and the corresponding citation impact scores can be found in Table 1 (for some sample papers) and in Table 3 and Table 4 (all sonified papers, in the appendix).

The introductory part is followed by a short description of the publications from the beginning of Loet’s career: “The 32 early papers in our dataset published by Loet appeared between 1996 and 2002 mostly in the journals *Scientometrics*, *Research Policy*, and *Journal of the American Society for Information Science and Technology*. Several papers in the early phase focus on a theory of citations. Although citations have been used in scientometrics as impact indicator since the 1970s, a generally accepted theory of citations has not been developed until today. Loet emphasized in this context the relevance of the social systems theory introduced by Niklas Luhmann. Niklas Luhmann is an influential and popular German sociologist. His theory is especially interesting for the scientometrics area, since the science system is conceptualized as a communication-based system and not a human-based system, and scientometricians use with publication and citation data formal communications as basis for their analyses. Besides papers on a theory of citations, the most influential paper published by Loet also falls into his early phase as a researcher. Together with Henry Etzkowitz, he proposed the Triple Helix of university-industry-government relations as a new model for explaining the modern science system".

After the spoken audio with the explanations of Loet's publications from the early phase, the following sonified metadata of publications are introduced with this sentence: "The sonification addresses three properties, first, single publication events over time, second, being open access or not, and third, field-normalized citation impact".

After the sonification of Loet's publications from the early phase, a short description of the publications from the end of Loet's career follows: "The 32 papers from the late phase were published between 2018 and 2020 in journals such as *Scientometrics*, *Journal of Informetrics*, and *Journal of the American Society for Information Science and Technology*. The papers deal with a diverse set of empirically and theoretically related topics. Some of the studies are based on the bibliometric method called reference publication year spectroscopy. The method can be used to analyze the historical roots of researchers, research areas, and research topics. Loet and co-authors applied the method to reveal the historical roots of Judith Bar-Ilan who was an influential information scientist and passed away in 2019. Other papers from the late phase of Loet focus on the integrated impact indicator. With some colleagues, he developed this indicator as a non-parametric alternative to established indicators in scientometrics. The indicator weights citation impact in accordance with the percentile rank class of each paper in a set of papers".

After the spoken audio with the explanations of Loet's publications from the late phase, it follows the sonified metadata of publications from the phase with this introducing sentence: "The sonification addresses three properties, first, single publication events over time, second, being open access or not, and third, field-normalized citation impact".

Our track ends with the following conclusions emphasizing the excellent past contributions of Loet to the field of scientometrics: "Loet was very interested in many research topics being they theoretically or empirically oriented. In research collaborations, we experienced him as a brilliant researcher who was enthusiastic in learning new approaches, methods or techniques. It was a pleasure to work with Loet, he always had excellent ideas and a fundamental background in science of science issues. But his contributions did not just end with research that led to publications; he also provided valuable software including source code on his webpage. His many contributions from a long career will surely continue to be very helpful to the scientometric community in the future".

### 4.2 Parameter mapping: the quantitative approach of sonification

The core of sonification is usually the parameter mapping part in the track. Whereas the spoken audio is a direct way of transporting information, the composition from parameter mapping is a symbolic approach: Variables are represented by sounds, whereas "the sounds are not directly generated out of the data, as with audification; only certain parameters of the sound event (loudness, pitch, timbre, duration, for example) are determined by characteristics of the data" (Dayé & de Campo, 2006, p. 355). The input in parameter mapping is certain data (e.g., citation impact or OA status of publications) and the output is parameters of sound events. Input and output are connected by a mapping schema explaining the connection between both (Lindborg et al., 2023). For composing the sonification of Loet's publications, we oriented our procedure towards previous sonifications explained in the literature (e.g., Agné et al., 2020; Larsen & Gilbert, 2013). The examples on the Loud Numbers website – designed by a team of specialists who help brands and organizations to create music based on data (see https://www.loudnumbers.net) – were also very inspiring.

For the sonification, we sorted Loet's publications by publication year and month in ascending order. The parameter mapping includes three properties of Loet's publications: (1) publication output, (2) publication appeared OA, and (3) received citation impact of a publication. The sounds representing the three properties are explained in the composed track in detail before the parameter mapping part starts later on in the track. We tried to select sounds for the track that listeners may intuitively associate with the event: "In sonification it matters which specific sound dimension is chosen to represent a given data dimension. This is partly because there seems to be some agreement among listeners about what sound attributes are good (or poor) at representing particular data dimensions" (Walker & Nees, 2011, p. 23).

For a paper published by Loet in a certain month, we selected a noise for the track that is similar to the noise frequently used by email clients: a wiping voice reflecting the sending of an email (ESM_Explainer_Video_One_Shot_Motion_Whoosh_Low_Air_Short_3.wav). Although an email is a short text compared to a scientific paper, it is similarly sent out as a paper with the intention to be read by a receiver. A deep bass tone is used to signal a publication that is openly available (OA paper, KSHMR_Sub_Drop_03.wav). We selected this deep bass tone to express "deep impact", i.e., a stronger impact and perception effect. The noise for publishing a certain paper (OA or not) starts a small sequence in the track that ends with the sonification of the citation impact (MNCS) of this paper.

Table 1. Schema for mapping field-normalized citation impact scores (MNCS)

| Lower bound | Upper bound | Meaning | Class | Tone F minor |
|---|---|---|---|---|
| | <=0.2 | Far below | 1 | F |
| 0.3 | 0.7 | Below | 2 | G |
| 0.8 | 1.2 | Average | 3 | G# |
| 1.3 | 1.6 | Above | 4 | A# |
| 1.7 | 2.2 | Far above | 5 | C |
| 2.3 | 4 | Outreaching | 6 | C# |
| >=4.1 | | Far outreaching | 7 | D# |

The pitch variation of a certain tone [MPE Erosion (Pressure) – Brass Quartet] in addition to an Auto-Filter (Cut-O-Move H) was used to reflect the MNCS of a paper following the approach of Agné et al. (2020). Table 1 shows how the impact parameters were mapped into tone whereby the schema follows the F minor key. The track (the sonified data) was composed therefore in F minor (F, G, G#, A#, C, C#, and D#). We selected the minor key for the track since the occasion for the sonification is a bereavement. Agné et al. (2020) argue that minor emphasizes the negative side of life (whereas major keys may be stronger associated with the positive side). Since it was not possible to assign a certain tone to the MNCS of every single paper, we formed seven citation impact classes – one class for each tone in the F minor scale following published guidelines for interpreting MNCS (Masterton & Sjödin, 2020; van Raan, 2004). These guidelines classify MNCS into several impact classes to support the interpretation of the scores.

The schema in Table 1, which connects MNCS and tone was used to sonify the 64 papers published by Loet in the early and late phase (the complete set of the mapped data can be found in Table 3 and Table 4 in the appendix). The resulting pitch variation of the tone depending on the MNCS is demonstrated in Table 2 for some example papers. The publication with the UT (identification number) "WOS:A1996UQ23300009" and an MNCS of 0.8, which is around an average citation impact, received the tone G#. From the outreaching impact of "WOS:A1997WW90100004" follows the D# tone, which is far away from G# (reflecting an average citation impact). The examples in the table demonstrate that the sonification is especially suitable for comparisons (Walker & Nees, 2011). The listener can not only compare the pitch variation of single publications, but also the structure of the parameter mappings resulting from the early and late phases' publications. For example, the listening comparison of both phases demonstrates that Loet published significantly more papers OA in the late than in the early phase.

Table 2. Field-normalized citation impact scores (MNCS) and OA status for some papers published by Loet

| UT | MNCS | OA status | Pitch variation |
|---|---|---|---|
| WOS:A1996UQ23300009 | 0.8 | | G# |
| WOS:A1996UX23900006 | 1.3 | | A# |
| WOS:A1996UM12100003 | 9.2 | | D# |
| WOS:A1997WP65300011 | 4.7 | 1 | D# |
| WOS:A1997XJ93700002 | 0.4 | 1 | G |
| WOS:A1997WW90100004 | 17.4 | | D# |
| WOS:A1997YG54300006 | 0.1 | | F |

### 4.3 The inclusion of electronic music

Although the sonification (parameter mapping) of data uses elements from music, one can question that sonification is music. Many sonifications that can be found in the Data Sonification Archive (see https://sonification.design) cannot be denoted as a musical or aesthetic experience (in our opinion). According to Vickers (2016), "music and sonification have ostensibly different goals. The composer strives for aesthetic interest, that is, the results should be 'aesthetically useful' … In sonification, it is not aesthetic interest but successful signification of the data that is the goal".

Although the goals between music and sonification may be different, many tracks have been published as mixes between music and sonification: "researchers publish papers describing their musical sonifications and composers present their sonification music" (Vickers, 2016). For example, sonification concerts were substantial parts of the ICAD meetings. One important reason for the incorporation of musical elements in sonifications may be "to engage and hold the listener's interest, surely a sonification that is more musical will be better than one that is not" (Vickers, 2016). For Walker and Nees (2011), however, it is important that the intended message of the sonification does not get lost with the consideration of music: “Although the resolution of issues regarding aesthetics and musicality is clearly relevant, it nevertheless remains advisable to design aesthetically pleasing (i.e., musical, etc.) sonifications to the extent possible while still conveying the intended message” (p. 27). The exchange of information should rank in front of the musicality (Vickers, 2016).

Especially the consideration of musical elements in science sonification may produce tracks that are able to communicate research results to people not working in the science sector. In the era of post-academic science, society (the government) expects these transfers from science (Bornmann, 2013; Tahamtan & Bornmann, 2020). The transfers may happen not only in the form of blogs, interviews or advice texts, but also in the form of sonifications (including musical elements). For example, Larsen and Gilbert (2013) drew inspiration from bebop jazz to produce “aesthetic musical compositions that interpret the relationships between elements in large biological datasets“ (Larsen & Gilbert, 2013). Other examples are mentioned by Walker and Nees (2011) and Vickers (2016) such as the concert “Global Music – The World by Ear” (Childs, 2007) and the sonification of solar wind data from NASA’s ACE satellite (Alexander, Zurbuchen, Gilbert, Lepri, & Raines, 2010). These sonifications are good examples for using sonifications as techniques to transfer research to people beyond science.

The sonification we produced based on Loet’s publications also includes musical elements. We asked Hennadii Boichenko to compose and play a piano cadence in F minor to reflect the sad occasion of the sonification. We expect that listeners will be put in the right mood for the occasion, as a minor cadence usually reflects a sad but flowing movement.

# 5 Discussion

Over the course of the centuries, visualizations have reached the status of the predominant method to present empirical results in science and beyond (Dayé & de Campo, 2006). Bibliometrics is no exception. We propose in this study metrics sonification as a new form of presenting bibliometric information that may develop to an alternative to metrics visualization or can be a useful addition to metrics visualizations. According to Dayé and de Campo (2006), with the introduction of sonification techniques in research, it is not intended to use “sonification as a replacement for visualisation. The thoughtful use of all the human senses, making good use of the perceptual strengths of each, is a more complex, but ultimately more fruitful undertaking. Where possible, it is advisable to combine both modes of perceptual representation, vision and hearing” (p. 353).

Other than visualizations, it is (still) unusual for interested scientists (people) in empirical research results to be confronted with sonifications. Today, one can expect that only a few people are able to interpret sonifications as desired: “We learn how to read graphical displays, and scientists’ skills are often highly developed and sophisticated in this area. But we do not learn how to identify structures or patterns in a given sequence of sounds” (Dayé & de Campo, 2006, p. 353). For a better understanding of sonifications, specific exercises are necessary requirements or the presentation includes visual and/or explaining elements (spoken audio) – as we did in this study.

In this paper, we present a possible application of metrics sonification in bibliometrics: the technique can be used to provide an insight in the career of an eminent scientist. We used the publications published by Loet as an example – Loet is a giant in the field of scientometrics who recently passed away. In the science of science area, his oeuvre was especially suitable for sonification, since Loet published many papers over his long career, and he published the papers in a broad range of topics. The sonification that we composed based on Loet’s oeuvre includes spoken audio, parameter mappings, and musical elements. We composed the track in F minor. The F minor parameter mapping and the piano cadence by Hennadii Boichenko in F minor as musical elements were chosen to express the sad occasion. It is a specific characteristic of sonifications that they are able to transport emotions very well (Diaz Gonçalves, 2023; Tulilaulu et al., 2018). This is scarcely possible with visualizations.

The sonification of papers published by an eminent scientist is only one possible scenario for the application of metrics sonification. It has been demonstrated in other disciplines than bibliometrics that many other applications are possible (and useful). One possible further application in bibliometrics may be the sonification of large publication sets (this case study focuses on only 64 publications). We can imagine that the sonification of the properties of large publication sets may help to understand the structure of the dataset and may facilitate comparisons of different large datasets. For example, we assume that the sonification of Loet's publications in the current study demonstrates already for untrained listeners that Loet significantly increased his number of OA publications from the early to the late phase of his career.

Another possible application of metrics sonification may be the accompaniment of visualizations. Visualizations are very popular in bibliometrics (see Figure 1), and the understanding of visualizations may be improved by sonifications. For example, the visualization of Loet's publications in Figure 1 demonstrates that they contain elements that are not all visible (but lie on top of each other). If the observer of such a network is interested in a certain element *i* in the visualization (e.g., the keyword "triple-helix"), a line could move across the network (similar to the moving line visible on radar monitors). A specific sound could be played always in that case when the moving line hits the element *i* (e.g. the keyword "triple-helix") on the network. We assume that such a combination of sonification and visualization can be used to find new or specific information, patterns, and relations in datasets (see Supper, 2015). Techniques which have been developed for the sonification of datasets including a spatial component may be helpful here. An overview of corresponding studies from this area can be found in Nasir and Roberts (2007).

Future applications of metrics sonifications will (probably) reveal which potential lies in this new technique for presenting (bibliometric) data. We recommend that metrics sonifications should be developed in multi-disciplinary teams to profit from different expertise and perspectives. Vickers (2016) sees "a value in a multidisciplinary approach, and this can work both ways: composers can help sonification designers, and sonification designers can help composers understand how data can be structured and mapped in interesting ways". The sonification team has to manage the task to integrate two expectations: data coverage and musical aesthetics. On the one hand, with the sonification, the team will be active in the area of music (Worrall, 2019), and aesthetic criteria should play an important role (to draw and hold the interest of listeners): "sound may aesthetically enhance a listener's interaction with a system" (Walker & Nees, 2011, p. 27). On the other hand, it is the task of the team to compose experimental sound that is especially influenced "by extra-subjective factors, by 'nature'" (Dayé & de Campo, 2006, p. 351). The team should avoid the risk that "the goal of communicating essential information … [is] masked in the effort to achieve a stronger musical expression" (Vickers, 2016). For a given sonification project, therefore, the team should try to find the appropriate balance between aesthetic preferences and the communication of information.

For the sonification of data, it is an important requirement that experts are involved who are able to understand and interpret not only the data itself, but also the statistics or methods that have produced the data (Worrall, 2019). Supper (2015) points out that the absence of these experts was a disadvantage in many previous sonifications: "From its very beginnings, the sonification community has focused on tools (often from the world of electronic music) and techniques for sonifying data rather than on the content and interpretation of the sonified data. This focus is an obstacle to the involvement of so-called domain scientists – specialists in the scientific domain from which data are sonified – within the sonification community. In the eyes of many sonification researchers, their (near) absence presents a major problem for the field. After all, for sonification to become successful, it is not enough that well-designed sonifications are made; they also have to be heard by specialists in different scientific fields: not just once out of curiosity, but repeatedly as part of their standard data analysis routines” (p. 458).

One of the most important advantages of data (metrics) sonification possibly lays in the combination and synergy of two areas: science and arts. While science communication is characterized by texts, numbers, and formulas, artists use a broad spectrum of channels: for example, “movement, images, sound, and sculpture” (Lindborg et al., 2023). For Agné et al. (2020), “synergies of arts and sciences are increasingly recognized for their potential to encourage intuitive and creative thinking, and therefore also as helpful for increasing public literacy and social engagement with political issues” (p. 272). The application of many channels for the communication of empirical data allows “users to hear, see, explore, and intuitively understand large amounts of information in as short a period of time, and with the minimum cognitive load possible” (Worrall, 2019, p. 255).

For identifying possible advantages of data sonification (in comparison to data visualization), research on data (and later metrics) sonification is necessary. Sonification projects should not be “dominated by … theoretical, arbitrary sound design choices“ (Walker & Nees, 2011, p. 22). Instead, on the one hand, sonification research should be used to explore new path routes for understanding and interpreting empirical research data (Dayé & de Campo, 2006). On the other hand, sonification research should run experiments to investigate whether the intended objectives of the sonification (e.g., the understanding of specific structures in a dataset by the listener) have been reached (Worrall, 2019). According to Lindborg et al. (2023), data sonification can reach the status of a scientific method only then if it fulfils the following scientific evaluation criteria that are tested in empirical studies on composed sonifications: systematicity, objectivity, and replicability. The authors have the impression that these criteria have not been fulfilled in many previous applications of data sonification.

More than a decade ago, (i) Dayé and de Campo (2006) and (ii) Kramer et al. (2010) assessed the previous research in the area of data sonification as follows: (i) "A significant effort has been made in the past ten to fifteen years to investigate the advantages of sonification, and to explore fields of application" (Dayé & de Campo, 2006, p. 354). (ii) "A rich history of research has provided valuable insight into the physiological, perceptual, and cognitive aspects of auditory perception for speech and relatively simple auditory events, such as pure tones and noise bursts" (Kramer et al., 2010, p. 6).

For Kramer et al. (2010) previous research has shown the following overarching results: "First, auditory perception is particularly sensitive to temporal characteristics, or changes in sounds over time. Human hearing is well designed to discriminate between periodic and aperiodic events and can detect small changes in the frequency of continuous signals. This points to a distinct advantage of auditory over visual displays … sonification is likely useful for comprehending or monitoring complex temporal data, or data that is embedded in other, more static, signals" (p. 6). Despite the great research efforts on data visualizations in previous years, Dayé and de Campo (2006) also confirm that "sonification still lacks a common methodological structure, an 'applied theory of sonification' to unify sonification research" (p. 355). Thus, there seems to be much room for improvements in research on data sonification (Zanella et al., 2022). This research should especially focus on experimental approaches that test and validate listener experiences (Lenzi & Ciuccarelli, 2020). Since sonifications are composed for listeners, their experiences with the sonifications should be in the focus of interest.

Based on the manifold experiences from fields other than bibliometrics, future metrics sonification projects should be accompanied by empirical evaluations.

## Acknowledgements

The bibliometric data used in this study are from a bibliometric in-house database of the Max Planck Society (MPG), developed and maintained in cooperation with the Max Planck Digital Library (MPDL, Munich), derived from the Science Citation Index Expanded (SCI-E), Social Sciences Citation Index (SSCI), and Arts and Humanities Citation Index (AHCI) prepared by Clarivate (Philadelphia, PA, USA) via the "Kompetenznetzwerk Bibliometrie" (see https://bibliometrie.info/en/about-kb/) funded by BMBF (grant 16WIK2101A). We are very grateful to Justin Salisbury for speaking up the spoken English audio and to Hennadii Boichenko to compose and play the piece "Minor Cadenza" for the track. We thank Miriam Quick (https://miriamquick.com) and Sven Hug for very valuable feedback to earlier versions of our metrics sonification. The available talks, articles, statements, and many other things which are available at https://www.loudnumbers.net/talks-articles were very inspiring material for composing the sonification and writing the paper.

## Competing interests

Lutz Bornmann is Editorial Board Member of *Scientometrics*. No funding was received for writing this paper. The authors declare they have no financial interests.

## Appendix

Table 3. Field-normalized citation impact scores (MNCS), OA status (1=yes), and pitch variation for papers published by Loet in early years

| Clarivate's accession number | MNCS | OA | Pitch |
|---|---|---|---|
| WOS:A1996UQ23300009 | 0.8 | | G# |
| WOS:A1996UX23900006 | 1.3 | | A# |
| WOS:A1996UM12100003 | 9.2 | | D# |
| WOS:A1997WP65300011 | 4.7 | 1 | D# |
| WOS:A1997XJ93700002 | 0.4 | 1 | G |
| WOS:A1997WW90100004 | 17.4 | | D# |
| WOS:A1997YG54300006 | 0.1 | | F |
| WOS:000071636000002 | 0.2 | | F |
| WOS:000075877700002 | 21.8 | | D# |
| WOS:000078882500003 | 2.3 | | C# |
| WOS:000079536100003 | 1.1 | 1 | G# |
| WOS:000089449100007 | 2.8 | | C# |
| WOS:000085125700001 | 55.0 | | D# |
| WOS:000085125700010 | 3.5 | | C# |
| WOS:000086051800001 | 1.1 | 1 | G# |
| WOS:000087404200005 | 1.4 | | A# |
| WOS:000088919200004 | 0.4 | | G |
| WOS:000166317500004 | 1.6 | 1 | A# |
| WOS:000167664900002 | 0.9 | | G# |
| WOS:000170034100002 | 0.5 | | G |
| WOS:000172450000009 | 1.8 | | C |
| WOS:000174044500009 | 2.1 | | C |
| WOS:000174373000007 | 0.6 | | G |
| WOS:000177967300003 | 2.4 | | C# |
| WOS:000179485100005 | 0.2 | | F |
| WOS:000180750800008 | 0.2 | | F |
| WOS:000186047400001 | 57.7 | | D# |
| WOS:000186047400014 | 8.7 | | D# |
| WOS:000223282100007 | 6.4 | | D# |
| WOS:000229836000001 | 0.9 | | G# |
| WOS:000228634800012 | 5.7 | 1 | D# |
| WOS:000229005500010 | 7.8 | 1 | D# |

Table 4. Field-normalized citation impact scores (MNCS), OA status (1=yes), and pitch variation for papers published by Loet in late years

| Clarivate's accession number | MNCS | OA | Pitch |
|---|---|---|---|
| WOS:000411017000010 | 0.3 | 1 | G |
| WOS:000406497200021 | 3.3 | 1 | C# |
| WOS:000424685100007 | 2.1 | 1 | C |
| WOS:000424685100016 | 6.0 | 1 | D# |
| WOS:000428351800056 | 0.3 | 1 | G |
| WOS:000429006100028 | 25.5 | 1 | D# |
| WOS:000438126800031 | 1.1 | 1 | G# |
| WOS:000442670600024 | 1.6 | 1 | A# |
| WOS:000442007200033 | 3.0 | 1 | C# |
| WOS:000445851400001 | 0.5 | 1 | G |
| WOS:000449961800009 | 1.9 | 1 | C |
| WOS:000451074800017 | 1.2 | | G# |
| WOS:000457778700009 | 1.3 | 1 | A# |
| WOS:000454956400009 | 4.2 | 1 | D# |
| WOS:000461544100001 | 0.6 | 1 | G |
| WOS:000460550800030 | 10.3 | 1 | D# |
| WOS:000463092700002 | 0.5 | 1 | G |
| WOS:000460804500022 | 4.5 | 1 | D# |
| WOS:000477740700019 | 0.2 | 1 | F |
| WOS:000469932800017 | 5.2 | 1 | D# |
| WOS:000464901100032 | 1.1 | 1 | G# |
| WOS:000463368300008 | 0.5 | 1 | G |
| WOS:000469058000017 | 1.8 | 1 | C |
| WOS:000493051600006 | 1.1 | | G# |
| WOS:000502883400008 | 0.7 | | G |
| WOS:000619261100002 | 7.0 | 1 | D# |
| WOS:000540811400007 | 2.7 | | C# |
| WOS:000528948000015 | 0.0 | 1 | F |
| WOS:000530786700001 | 0.7 | 1 | G |
| WOS:000530786700010 | 1.0 | 1 | G# |
| WOS:000523401300003 | 0.0 | | F |
| WOS:000595260100016 | 0.0 | 1 | F |